\documentclass[preprint,superscriptaddress,amsmath,amssymb,prd,aps,showpacs,floatfix,nofootinbib]{revtex4-1}
\usepackage{graphicx}
\usepackage{dcolumn}
\usepackage{bm}
\usepackage{mathrsfs}
\usepackage{hyperref}
\usepackage{amsmath}
\usepackage{amssymb}
\usepackage{bm}
\usepackage{color}
\usepackage{float}
\usepackage{dcolumn}
\usepackage{multirow}
\usepackage{changepage}
\usepackage{enumerate}
\usepackage[normalem]{ulem}
\usepackage[dvipsnames]{xcolor}
\usepackage{braket}
\usepackage{nicefrac,ulem}
\usepackage{hyperref}
\hypersetup{
    colorlinks=true,
    urlcolor=blue,
    linkcolor = blue,
    urlcolor  = blue,
    citecolor = blue,
    anchorcolor = blue,
    pdftitle={a1 to pi sigma},
    pdfauthor={friends},
    pdfsubject={a1}}

\newcommand{\mev}{\textrm{ MeV}}

\usepackage[commentmarkup=todo,authormarkup=superscript,todonotes={textsize=tiny}]{changes}

\begin{document}
\title{The $\pi f_0(500)$ decay of the $a_1(1260)$}
\date{\today}

\author{R.~Molina}
\email{Raquel.Molina@ific.uv.es}
\affiliation{Departamento de F\'{\i}sica Te\'orica and IFIC, Centro Mixto Universidad de Valencia-CSIC Institutos de Investigaci\'on de Paterna, Aptdo.22085, 46071 Valencia, Spain}

\author{M.~D\"oring}
\email{doring@gwu.edu}
\affiliation{Department of Physics, The George Washington University, Washington, DC 20052, USA}
\affiliation{Theory Center, Thomas Jefferson National Accelerator Facility, Newport News, Virginia 23606, USA}

\author{W.~H.~Liang}
\email{liangwh@gxnu.edu.cn}
\affiliation{Department of Physics, Guangxi Normal University, Guilin 541004, China}
\affiliation{Guangxi Key Laboratory of Nuclear Physics and Technology, Guangxi Normal University, Guilin 541004, China}

\author{E.~Oset}
\email{oset@ific.uv.es}
\affiliation{Department of Physics, Guangxi Normal University, Guilin 541004, China}
\affiliation{Departamento de F\'{\i}sica Te\'orica and IFIC, Centro Mixto Universidad de
Valencia-CSIC Institutos de Investigaci\'on de Paterna, Aptdo.22085,
46071 Valencia, Spain}

\preprint{JLAB-THY-21-3464}

\begin{abstract}
  We evaluate the $a_1(1260) \to \pi \sigma (f_0(500))$ decay width from the perspective that the $a_1(1260)$ resonance is dynamically generated from the pseudoscalar-vector interaction and the $\sigma$ arises from the pseudoscalar-pseudoscalar interaction. A triangle mechanism with $a_1(1260) \to \rho \pi$ followed by $\rho \to \pi \pi$ and a fusion of two pions within the loop to produce the $\sigma$ provides the mechanism for this decay under these assumptions for the nature of the two resonances. We obtain widths of the order of $13-22$ MeV. Present experimental results  differ substantially from each other, suggesting that extra efforts should be devoted to the precise extraction of this important partial decay width, which should provide valuable information on the nature of the axial vector and scalar meson resonances and help clarify the role of the $\pi\sigma$ channel in recent lattice QCD calculations of the $a_1$. 
\end{abstract}

\maketitle

\section{Introduction}
\label{sec:intro}

The nature of the axial vector meson resonances is a subject of continuous debate.
It is well known that quark models, very successful in correlating a vast amount of data concerning meson spectra,
have difficulties reproducing the properties of axial vector mesons \cite{Isgur,Vijande}.
On the other hand, the advent of the chiral unitary approach,
combining chiral dynamics with a unitary approach in coupled channels,
showed that the lowest-energy axial vector mesons can be fairly well reproduced from the interaction of vector mesons with pseudoscalars \cite{Lutz,Luis,GengRoca,RenGeng}.
The interaction used between vector-pseudoscalar ($VP$) channels is taken from the chiral Lagrangian of Ref.~\cite{Birse}.
In Ref.~\cite{Nagahiro}, for the particular case of the $\rho \pi$ interaction, and in Ref.~\cite{Jorgivan} for the general $VP$ interaction,
it was shown that this Lagrangian can be obtained from the exchange of vector mesons within the local hidden gauge approach \cite{hidden1,hidden2,hidden4,Nagahiro}.
This conclusion is in line with the findings in the pseudoscalar-pseudoscalar ($PP$) interaction \cite{Rafael}.
However,the $PV$ interaction shows a peculiarity
since one can make the transition $PV \to VP$ mediated by a pseudoscalar exchange of one $P$ from the decay of $V \to PP$.
Fortunately, such contributions are very small compared to those from vector exchange,
as recently shown in Ref.~\cite{Jorgivan},
something corroborated in the study of $D$ decays into three mesons looking at meson final state interactions \cite{Nakamura}.
Thus, one can rely on the information for the axial vector mesons obtained within the chiral unitary approach
with the chiral Lagrangian of Ref.~\cite{Birse}.
Lattice QCD has also shown interest in describing the axial vector resonances.
Concretely, the $a_1(1260)$ was already studied in Ref.~\cite{Sasa} (see Ref.~\cite{RocaFinite} in connection to this work).
More detailed work on the $a_1(1260)$ and $b_1(1235)$ is shown in Ref.~\cite{Lang},
and the $b_1(1235)$ is also investigated along this line considering the coupled channels $\pi \omega, \pi \phi$ in Ref.~\cite{Dudek}.
The first extraction of a three-body resonance from lattice QCD, using up to three pion operators and a suitable three-body finite-volume formalism was achieved recently for the example of the $a_1(1260)$~\cite{Mai:2021nul}. The finite-volume mapping (FVU) relies on unitarity for three particles~\cite{Mai:2017vot} leading to the aforementioned pseudoscalar exchange which develops singularities in the finite volume~\cite{Mai:2017bge}. The largest source of systematics in the $a_1(1260)$ pole extraction from lattice QCD is the role of the $\pi\sigma$ channel~\cite{Mai:2021nul}. The prediction of the partial decay width $a_1\to\pi\sigma$ in the present paper is, therefore, essential to assess the role of such sub-dominant channels in upcoming lattice QCD calculations.

As these examples show, combined work using effective field theories, quark models, lattice QCD simulations
and experiment will ultimately unveil the nature of the axial vector meson resonances,
and work goes along in these direction.
One of the successful strategies pursuing this aim is to look for particular decay modes of these states
which reveal the coupling of the resonances to some bound meson-meson components.
Let us take as an example the $f_1(1285)$, which appears as a $K^* \bar K +c.c.$ single-channel state, bound by about $100 \mev$ \cite{Lutz,Luis,RenGeng}.
The $f_1(1285)$ can naturally decay to $K^*\bar K$ (due to the $K^*$ width),
then the $K^*$ decays to $K\pi$ and $K\bar K$ fuses into $a_0(980)$, which couples strongly to $K\bar K$.
This process constitutes a triangle mechanism that leads to the $f_1(1285) \to \pi a_0$ decay
which is tied to the coupling of the $f_1(1285)$ to the bound $K^*\bar K$ component.
This work is done in Ref.~\cite{Aceti} and a good agreement with experiment is found.
In a similar way one can study the $f_1(1285) \to \pi K\bar K$ decay,
and a good agreement with experiment is found in Ref.~\cite{AcetiKK}.

The triangle mechanisms have been widely used to test molecular components, with the information encoded in the resonance decay vertex of the first step, 
or on the fusion vertex of the last step.
In Refs.~\cite{Zou,Acetieta} a triangle mechanism was used to explain the unexpectedly large isospin breaking of the $\eta(1405) \to \pi^0 f_0(980)$ decay
compared to $\eta(1405) \to \pi^0 a_0(980)$ observed at BES \cite{BESeta} (see also following works of Refs.~\cite{Achasov,QiangZhao}).
Recent work on triangle mechanisms to describe different processes can be seen in Refs.~\cite{He,HuangGeng,XieChi,Xiemore,HuangXie,YuKe,GengXie}.

One particular case of a triangle mechanisms is the one that develops a triangle singularity (TS).
This was early discussed in Refs.~\cite{Karplus,Landau}, which showed
that a singularity in the triangle diagram emerged when the three particles in the loop are placed simultaneously on shell,
while being collinear and the decay of the first step and fusion of the second one
are possible at a classical level (Coleman-Norton theorem \cite{Norton}).
The topic has experienced a rebirth recently
since one finds nowadays many physical examples of reactions which are interpreted in terms of a TS.
A modern reformulation of the problem, intuitive and practical, is given in Ref.~\cite{BayarGuo}
and a recent review on the topic is given in Ref.~\cite{Guorev}.

Many examples of TS are discussed in Ref.~\cite{Guorev}.
Here we just mention two examples which have particular relevance.
One of them is the TS developed in the $a_1(1260)$ decay into $\pi f_0(980)$,
related to the topic that we face in this paper (the $a_1(1260) \to \pi \sigma (f_0(500))$ decay)
which was originally associated to a new resonance ``$a_1(1420)$" by the COMPASS Collaboration \cite{COMPASS},
but which was soon interpreted in terms of a TS \cite{Oka,Mikha,AceDai} and acknowledged as such by the COMPASS Collaboration \cite{Alexeev}.
The other recent example is the case of the $pp \to \pi^+ d$ fusion reaction,
which, as shown in Ref.~\cite{IkenoRaq}, has an unexpectedly large strength due to a TS.
This work, together with the one of Ref.~\cite{RaquelIke},
present an alternative explanation of the peak observed in the $pn \to \pi^0 \pi^0 d$ ($pn \to \pi^+ \pi^- d$) reaction,
so far associated to a dibaryon ``$d^*(2380)$" \cite{dibaryon}.

It is interesting to recall that not all triangle diagrams develop a TS,
actually these are exceptional cases,
but it is easy to test if a triangle diagram develops a TS applying a simple equation (see Eq.~(18) of Ref.~\cite{BayarGuo}).
We shall take advantage of this equation too,
because it not only serves to see if there is a TS,
but also to have a feeling of the strength of the loop by observing how far this equation is from being fulfilled.

In the present work we address the issue of the partial decay width of the $a_1(1260)$ into $\pi \sigma (f_0(500))$.
This can be seen for instance in the $a^+_1(1260) \to \pi^+ \pi^0 \pi^0$ decay,
where the $\pi^0 \pi^0$ will come from the $\sigma$ meson,
hence, this is part of the three pion decay of the $a_1(1260)$.
This latter problem has been extensively studied from different points of view and it is not our purpose to go through it again.
A thorough account and discussions on the different approaches can be seen in the works of Refs.~\cite{Misha,ZhangXie,DaiRoca},
but in none of those works the $\pi \sigma$ decay mode has been isolated.

Experimentally there are only a few experiments tabulated in the PDG \cite{PDG} (although ``they are not taken for averages, fits, limts, etc") and the branching ratios are
\begin{equation}
  \Gamma[a_1 \to \pi f_0(500), f_0 \to \pi \pi]/\Gamma_{\rm tot}=(18.76\pm 4.29 \pm 1.48) \times 10^{-2} ~\text{\cite{Asner}},\label{eq:asner}
\end{equation}
\begin{align}
\Gamma[a_1 \to \pi f_0(500), f_0 \to \pi \pi]/\Gamma[(\rho \pi)_{S{\rm{-wave}}}, \rho \to \pi \pi]=
 \left\{
  \begin{array}{l}
   ~~ (6\pm 5)\times 10^{-2}~\text{\cite{Salvini}}, \\
    \sim 30 \times 10^{-2}~\text{\cite{Akhme}}, \\
   ~~ (0.3\pm 0.3) \times 10^{-2}~\text{\cite{Longa}},
  \end{array}
 \right.
 \label{eq:exp}
\end{align}
and in Ref.~\cite{Chung} this decay mode is reported as ``seen".
As we can see,
the decay ratios are very different,
and with large uncertainties.
The data of Ref.~\cite{Asner} come from the $\tau^- \to \nu_\tau \pi^- \pi^0 \pi^0$ decay measured at CLEO II. These data should be taken with certain caution since $m_\sigma =860 \mev$ and $\Gamma_\sigma=880 \mev$ are used in the substructure fits, which are
very different from present values \cite{PDG,Colangelo,sigma}. \footnote{From Ref.~\cite{Colangelo} we find
\begin{equation*}
  m_\sigma =470 \pm 30 \mev,~~~~~\Gamma_\sigma=2\times (295\pm 20) \mev,
\end{equation*}
and from Ref.~\cite{sigma}
\begin{equation*}
  m_\sigma =449^{+22}_{-16} \mev,~~~~~\Gamma_\sigma=2\times (275\pm 12) \mev.
\end{equation*}
We take values close to these  as $m_\sigma=458$~MeV and $\Gamma_\sigma =464$~MeV, coming from the use of Eqs.~\eqref{eq:betheu}, \eqref{eq:potu}.
}
Dalitz plots with a more realistic $\sigma$ resonance were predicted by the three-body framework of Ref.~\cite{Kamano:2011ih}. That work and Ref.~\cite{Misha} use two-body amplitudes for the sub-channels in contrast to earlier work~\cite{Janssen:1993nj} in this direction. However, none of these frameworks or related theoretical studies~\cite{Wagner:2007wy, Wagner:2008gz} have performed a refit of the CLEO Dalitz distributions~\cite{Asner} which are among the most precise sources of information on the $a_1$ decay modes. The ALEPH data on the lineshape~\cite{ALEPH:2005qgp}, in contrast, served for the determination of the pole position of the $a_1$~\cite{Janssen:1993nj,Mikha,Mai:2021nul} but does not contain information on branching ratios.

In order to evaluate the $a_1(1260) \to \pi \sigma$ decay width,
we follow the procedure of Ref.~\cite{ZhangXie} in the study of $a_1(1260) \to \pi^+ \pi^+\pi^-$ decay,
where the tree level $a_1(1260) \to \rho \pi$ together with final state interaction of two pions via a triangle loop mechanism are considered.
The formalism follows closely the one of Refs.~\cite{Aceti,AceDai} and is based on the chiral unitary approach for the axial vector mesons \cite{Luis} and for scalar mesons \cite{npa}.
The present work has to be considered as complementary to the one of Ref.~\cite{ZhangXie},
where the $\pi^+ \pi^-$ mass distribution is obtained, that
is largely dominated by the $a_1 \to\rho \pi$ tree level mechanism,
and that is shown to be consistent with the experimental data of ARGUS \cite{ARGUS}.
In the low energy part of the $\pi^+ \pi^-$ invariant mass spectrum,
some extra strength appears which should be attributed to the $\pi^+ \sigma$ production but the $\pi \sigma$ decay mode was not isolated.
The precise evaluation of this decay mode is the purpose of the present work.

\section{Formalism}
\label{sec:form}
\begin{figure}
   \includegraphics[width=0.9\linewidth]{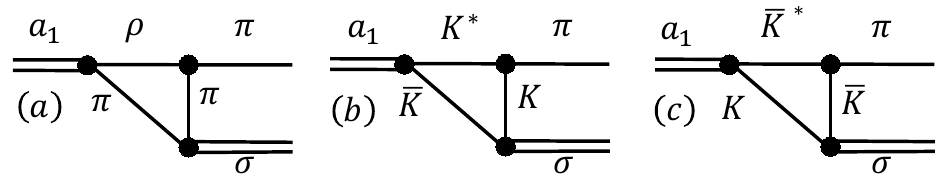}
    \caption{Diagrams leading to the $a_1(1260)\to\pi\sigma$ decay: (a) through $a_1\to\pi\rho$ in the first step; (b) through $a_1\to K^*\bar K$ in the first step; (c) through $a_1\to \bar K^*K$ in the first step. }
    \label{fig:triangles}
\end{figure}

We follow closely the approach of Ref.~\cite{Aceti} for the $f_1(1285)\to \pi a_0(980)$ decay. In that case the $f_1(1285)$ decays to $K^*\bar K$, the $K^*$ to $K\pi$ and then the $K\bar K$ fuse to give the $a_0(980)$. In the present case, we have the $a_1(1260)$ coupled to $\rho\pi$ and $\bar K^*K-K^*\bar K$, with the largest coupling to $\rho\pi$~\cite{Luis}. In view of this, and similar to Ref.~\cite{ZhangXie}, the mechanisms to produce the $\sigma$ are given by the diagrams of Fig.~\ref{fig:triangles}.
One can see that none of these triangle mechanisms produces a TS~\cite{BayarGuo}, yet the diagrams Fig. \ref{fig:triangles} (b) and Fig. \ref{fig:triangles} (c) are highly suppressed with respect to diagram Fig. \ref{fig:triangles} (a). Indeed, in diagram Fig. \ref{fig:triangles} (a) one can place the $\rho$, $\pi$, $\pi$ intermediate states on shell, yet the condition of the Coleman-Norton theorem~\cite{Norton} is not satisfied, since, in a collinear decay, the $\pi$ coming from the $\rho$ decay has a smaller velocity than the one from $a_1$ decay,  hence, it cannot catch up with the latter pion to produce the $\sigma$. The processes in Fig. \ref{fig:triangles} (b) and Fig. \ref{fig:triangles} (c) are much weaker since, first, the $a_1\to \bar K^*K$ coupling is smaller than the $a_1\to \rho\pi$ one~\cite{Luis}, second, because the $\sigma$ couples strongly to  $\pi\pi$  but very weakly to $K\bar K$~\cite{npa} and, third, because the $K\bar K$ state is highly off shell in the loop if we wish to produce the $\sigma$. Hence, we shall not consider diagrams Fig. \ref{fig:triangles} (b) and Fig. \ref{fig:triangles} (c) in our evaluation.

It is curious to note that, conversely, the diagrams Fig. \ref{fig:triangles} (b) and Fig. \ref{fig:triangles} (c) are those responsible for the $a_1(1260)\to\pi f_0(980)$ decay, replacing the $\sigma$ by the $f_0(980)$~\cite{AceDai}. This is because in this case the diagrams develop a TS, and also the $f_0(980)$ coupling to $K\bar K$ is much stronger than the coupling of the $\sigma$ to $K\bar{K}$~\cite{nsd}. This decay mode was the one originally assigned to the $a_1(1420)$ resonance, now accepted as a TS~\cite{Alexeev}. Yet, the contribution of this mode to the $a_1(1260)$ width is small, of the order of 1~MeV~\cite{AceDai} which gives us an idea of the smallness of the diagrams in Fig. \ref{fig:triangles} (b) and Fig. \ref{fig:triangles} (c) in the much less favored actual situation of $\sigma$ production.

\begin{figure}
   \includegraphics[width=0.85\linewidth]{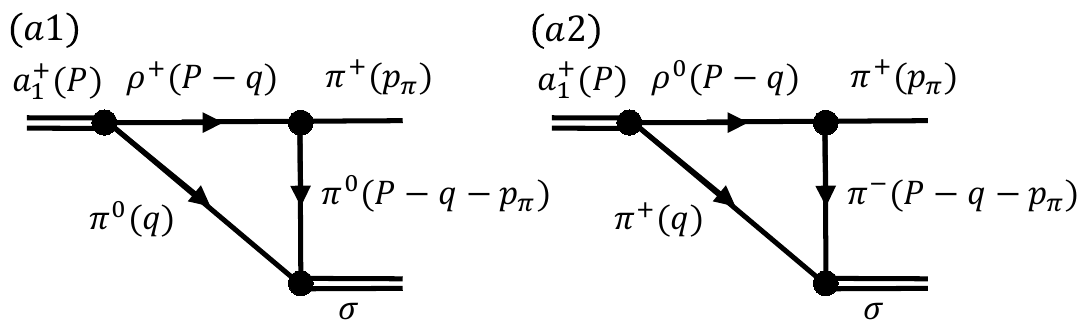}
    \caption{Detail of diagram (a) of Fig~\ref{fig:triangles} showing explicitly the $\rho^+\pi^0$ and $\rho^0\pi^+$ decays of the $a_1$. Particle momenta are shown in parentheses.}
    \label{fig:details}
\end{figure}

We, thus, proceed with the diagram of Fig.~\ref{fig:triangles}(a) to produce $\pi\sigma$ and we show in detail in Fig.~\ref{fig:details} the diagrams to consider. 
We need the states of the $\sigma$ and $a_1^+$ in terms of $\pi$ and $\rho$ multiplets $(-\pi^+,\pi^0,\pi^-)$, $(-\rho^+,\rho^0,\rho^-)$, 
\begin{align}
\ket{\pi\pi,\,I=0}&=\frac{-1}{\sqrt{3}}\ket{\pi^+\pi^-+\pi^-\pi^++\pi^0\pi^0} \ ,\label{eq:pipi}\\
\ket{\rho\pi,\,I=1,I_3=1}&=\frac{1}{\sqrt{2}}\ket{\rho^0\pi^+-\rho^+\pi^0} \ .
\end{align}
We take the coupling $g_{\sigma,\pi\pi}$ of the $\sigma$ to $\pi\pi$ from the chiral unitary approach \cite{npa} with a single channel $\pi\pi$ in $I=0$, but we also consider the $\pi\pi$, $I=0$ amplitudes in coupled channels. Further details will be given at the beginning of Sec.~\ref{sec:results} and Appendix~\ref{sec:appa}. In the chiral unitary approach, for convenience, one works with a unitary normalization where Eq. (\ref{eq:pipi}) has a coefficient $1/\sqrt{6}$ instead of $1/\sqrt{3}$. This $\sqrt{2}$ factor has to be restored at the end for the external pions so that
\begin{equation}
    g_{\sigma,\pi\pi}=\sqrt{2}\,(1248-i\,2717)~\mathrm{MeV}\ .\label{eq:cousigma}
\end{equation}
Similarly, from Ref.~\cite{Luis}, in the $a_1$ rest frame, the decay vertex reads
\begin{align}
-it_{a_1,\rho\pi}&=-ig_{a_1,\rho\pi}\vec\epsilon_{a_1}\cdot\vec\epsilon_\rho \ ,
\label{eq:a1pirho}
\end{align}
where $\vec\epsilon_{a_1},\,\vec\epsilon_\rho$ are the polarization vectors of the $a_1,\,\rho$, with 
\begin{equation}
    g_{a_1,\rho\pi}=(-3795+i\, 2330)~\mathrm{MeV}\ .
    \label{eq:ga1}
\end{equation}
Thus,
\begin{eqnarray}
 g_{a_1,\rho^0\pi^+}=\frac{1}{\sqrt{2}}g_{a_1,\rho\pi} \ ,\quad g_{a_1,\rho^+\pi^0}=-\frac{1}{\sqrt{2}}g_{a_1,\rho\pi} \ . 
\end{eqnarray}
 Analogously,  we have (see Appendix~\ref{sec:appa})
 \begin{eqnarray}
   g_{\sigma,\pi^+\pi^-}=\frac{1}{\sqrt{3}}g_{\sigma,\pi\pi} \ ,\quad g_{\sigma,\pi^0\pi^0}=\frac{1}{\sqrt{3}}g_{\sigma,\pi\pi} \ . \end{eqnarray}
 We also need the $\rho\to\pi\pi$ Lagrangian which is given by
\begin{align}
{\cal L}=-ig\braket{V^\mu\left[P,\partial_\mu P\right]}\ ,\quad g=\frac{M_V}{2f} \ , \quad (M_V\approx 800\text{ MeV, }
f=93\text{ MeV}) \ ,
\end{align}
with $V,P$ the SU(3) matrices for the vector mesons and pseudoscalars, respectively~\cite{Luis}, which gives
\begin{align}
-it_{\rho^+,\pi^+\pi^0}&=i\sqrt{2}g\left(2\vec p_\pi+\vec q\,\right)\cdot\vec\epsilon_\rho \ ,\nonumber \\
-it_{\rho^0,\pi^+\pi^-}&=-i\sqrt{2}g\left(2\vec p_\pi+\vec q\,\right)\cdot\vec\epsilon_\rho \ ,
\end{align}
where we keep only the spatial components of the $\rho$ polarization, given the coupling of Eq.~\eqref{eq:a1pirho} and the fact that the $\epsilon^0$ component provides only a very small contribution even for $\rho$ mesons with the momenta involved in the $a_1\to\rho\pi$ decay (see Appendix~A of Ref.~\cite{SakaiRamos}).

With these ingredients we obtain the amplitudes corresponding to the diagrams of Fig.~\ref{fig:details} (a1) and (a2):
\begin{align}
-it^{(a1)}&=\int\frac{d^4 q}{(2\pi)^4}\,(-i)\,\left(-\frac{1}{\sqrt{2}}\right)\,g_{a_1,\rho\pi}\,\vec\epsilon_{a_1}\cdot\vec\epsilon_\rho\,ig\sqrt{2}\left(2\vec p_\pi+\vec q\,\right)\cdot\vec\epsilon_\rho\,(-i)\left(-\frac{1}{\sqrt{3}}\right)g_{\sigma,\pi\pi}\, \Pi(q)\ ,\label{eq:ta1}\\
-it^{(a2)}&=\int\frac{d^4 q}{(2\pi)^4}\,(-i)\,\left(\frac{1}{\sqrt{2}}\right)\,g_{a_1,\rho\pi}\,\vec\epsilon_{a_1}\cdot\vec\epsilon_\rho\,ig(-\sqrt{2})\left(2\vec p_\pi+\vec q\,\right)\cdot\vec\epsilon_\rho\,(-i)\left(-\frac{1}{\sqrt{3}}\right)g_{\sigma,\pi\pi}\, \Pi(q)\ ,
\label{eq:amplitudes}
\end{align}
where $\Pi$ is the product of propagators
\begin{align}
\Pi(q)&=\frac{1}{2\omega_\rho}\,\frac{i}{P^0-q^0-\omega_\rho+\frac{i\Gamma_\rho}{2}}\,\frac{i}{q^2-m_\pi^2+i\epsilon}\,\frac{i}{(P-q-p_\pi)^2-m_\pi^2+i\epsilon} \  ,
\label{eq:propagators}
\end{align}
with $\omega_\rho^2=m_\rho^2+\vec q^{\,\,2}$. In Eq.~\eqref{eq:propagators} we have kept the positive-energy part of the $\rho$ propagator, corresponding to a heavy particle that will be mostly on shell in the loop, but keep the full propagators for the two pions in the loop. Separating the pion propagators into their positive and negative energy parts, 
\begin{align}
\frac{1}{q^2-m_\pi^2+i\epsilon}=\frac{1}{2\omega(\vec q\,)}\left(\frac{1}{q^0-\omega(\vec q\,)+i\epsilon}-\frac{1}{q^0+\omega(\vec q\,)-i\epsilon}\right),
\end{align}
where $\omega^2(\vec q\,)=\vec q^{\,\,2}+m_\pi^2$, we can readily perform the $q^0$ integration of the amplitudes $t^{(a1)}$ and $t^{(a_2)}$ of Eqs.~\eqref{eq:ta1} and \eqref{eq:amplitudes}. We obtain
\begin{align}
t&=t^{(a1)}+t^{(a2)}=\frac{-2}{\sqrt{3}}\,g\,g_{a_1,\rho\pi}\,g_{\sigma,\pi\pi}\,\epsilon_{a_1,j}\int\frac{d^3q}{(2\pi)^3}\,F(\vec q,\vec p_\pi)\,(2\vec p_\pi+\vec q\,)_j
\end{align}
where
\begin{align}
F(\vec q,\vec p_\pi)&=\frac{1}{2\omega(\vec q\,)}\,\frac{1}{2\omega_\rho}\,\frac{1}{2\omega (\vec q+\vec p_\pi)}\,\bigg[\frac{1}{P^0-\omega(\vec q\,)-\omega_\rho(\vec{q}\,)+\frac{i\Gamma_\rho}{2}}\nonumber \\
&\times\left(\frac{1}{P^0-p_\pi^0-\omega(\vec q\,)-\omega(\vec q+\vec p_\pi)+i\epsilon}+\frac{1}{p_\pi^0-\omega_\rho(\vec{q}\,)-\omega(\vec q+\vec p_\pi)+\frac{i\Gamma_\rho}{2}}\right)\nonumber \\
&+\frac{1}{p_\pi^0-\omega(\vec q+\vec p_\pi)-\omega_\rho(\vec{q}\,)+\frac{i\Gamma_\rho}{2}}\,\frac{1}{p_\pi^0-P^0-\omega(\vec q\,)-\omega(\vec q+\vec p_\pi)+i\epsilon}\bigg],
\end{align}
which, taking into account that 
\begin{align}
\int\frac{d^3q}{(2\pi)^3}\,F(\vec q,\vec p_\pi)\,q_i=p_{\pi,i}\int\frac{d^3q}{(2\pi)^3} F(\vec q,\vec p_\pi)\frac{\vec p_\pi\cdot\vec q}{\vec p_\pi^{\,\,2}} \ ,
\end{align}
can be cast as
\begin{align}
t&=\frac{-2}{\sqrt{3}}\,g\,g_{a_1,\rho\pi}\,g_{\sigma,\pi\pi}\,\vec\epsilon_{a_1}\cdot\vec p_\pi\int\frac{d^3q}{(2\pi)^3}\,F(\vec q,\vec p_\pi)\left(2+\frac{\vec p_\pi\cdot\vec q}{\vec p_\pi^{\,\,2}}\,\right) \ .
\label{eq:tready}
\end{align}
In Ref.~\cite{ZhangXie} the integral of Eq.~\eqref{eq:tready} is regularized in terms of a cut off in $\vec q$ of 630~MeV and an additional form factor. We regularize it with the intrinsic cut offs that stem from the chiral unitary approach. Indeed, as seen in Ref.~\cite{DaniJuan}, the scattering amplitudes appear with the factorization 
\begin{align}
t(\vec q,\vec q\,')=t\,\theta(q_\text{max}-|\vec q\,|)\,\theta(q'_\text{max}-|\vec q\,'\,|) \ ,
\end{align}
implying that the vertices are factorized as
\begin{align}
t_{a_1,\rho\pi}&\equiv g_{a_1,\rho\pi}\,\theta(q_\text{max}-|\vec q\,|)\,\vec\epsilon_{a_1}\cdot\vec\epsilon_\rho
\nonumber \\
t_{\sigma,\pi\pi}&\equiv g_{\sigma,\pi\pi}\,\theta(q'_\text{max}-|\vec q\,'|) \ ,
\nonumber
\end{align}
and $\vec q\,'$ is the boosted value of the momentum $\vec q$ to the frame where the $\sigma$ is at rest (see formula in Appendix~\ref{sec:appb}). Thus, finally,
\begin{align}
t=-\frac{2}{\sqrt{3}}\,g\, g_{a_1,\rho\pi}\,g_{\sigma\pi\pi}\,\vec\epsilon_{a_1}\cdot\vec p_\pi\,\tilde t
\end{align}
where 
\begin{align}
\tilde t&=\int\frac{d^3q}{(2\pi)^3}\,F(\vec q,\vec p_\pi)\left(2+\frac{\vec p_\pi\cdot \vec q}{p_\pi^{\,\,2}}\,\right)\,\theta(q_\text{max}-|\vec q\,|)\,\theta(q'_\text{max}-|\vec q\,'|)
\end{align}
and, following Ref.~\cite{Luis}, we take $q_\text{max}=800$~MeV and $q'_\text{max}=750$~MeV (as discussed in Sec.~\ref{sec:results}), by means of which we obtain the width for $a_1\to\pi\sigma$, averaging over the $a_1$ initial polarizations,
\begin{align}
\Gamma(M_I)&=\frac{1}{8\pi}\frac{1}{M_{a_1}^2}\frac{1}{3}\left|\frac{2}{\sqrt{3}}\,g\,g_{a_1,\rho\pi}\,g_{\sigma,\pi\pi}\right|^2\, p_\pi^3\,|\tilde t|^2
\label{eq:width}
\end{align}
where $M_I$ is the invariant mass of the $\sigma$ and $p_\pi\equiv|\vec p_\pi|=\lambda^{\nicefrac{1}{2}}\left(M_{a_1}^2,m_\pi^2,M_I^2\right)/\left(2M_{a_1}\right)$ here and in the following.
Eq.~\eqref{eq:width} assumes a $\sigma$ with fixed mass $M_I$, but the $\sigma$ has, in fact, a large mass distribution. We can formally take this into account (it will done more rigorously below) assuming a Breit-Wigner mass distribution
\begin{align}
S_\sigma(M_I)=-\frac{1}{\pi}\,\text{Im}\,\frac{1}{M_I^2-m_\sigma^2+i m_\sigma\Gamma_\sigma}
\end{align}
and, thus, 
\begin{align}
\Gamma=\frac{-1}{18\pi^2M_{a_1}^2}\,g^2\,|g_{a_1,\rho\pi}|^2\int dM_I^2\,\text{Im}\frac{\vert g_{\sigma,\pi\pi}\vert^2}{M_I^2-m_\sigma^2+i m_\sigma\Gamma_\sigma}
|\tilde t(M_I)|^2\,p_\pi^3(M_I) \ .
\label{eq:gammainter}
\end{align}
But now we see that formally $\vert g_{\sigma,\pi\pi}\vert^2(M_I^2-m_\sigma^2+i\,m_\sigma\Gamma_\sigma)^{-1}$ is the Breit-Wigner representation of the $\pi\pi$ amplitude. Hence, we replace, 
\begin{align}
\frac{\vert g_{\sigma,\pi\pi}\vert^2}{M_I^2-m_\sigma^2+im_\sigma\Gamma_\sigma}\to t_{\pi\pi,\pi\pi}^{I=0}(M_I) \ ,
\end{align}
which allows us to use the realistic scattering amplitude $t_{\pi\pi,\pi\pi}^{I=0}$ from Ref.~\cite{npa}, or from Ref.~\cite{Liang} where the coupled channels $\pi^+\pi^-$, $\pi^0\pi^0$, $K^+K^-$, $K^0\bar{K}^0$ and $\eta\eta$ are used. Then, as found in Appendix~\ref{sec:appa}, 
\begin{align}
t_{\pi\pi,\pi\pi}^{I=0}=3t_{\pi^+\pi^-,\pi^+\pi^-}^{I=0} \ .\label{eq:trel}
\end{align}
Hence, we can write 
\begin{align}
\Gamma(M_{a_1})=\frac{-1}{18\pi^2M_{a_1}^2}\,g^2\,|g_{a_1,\rho\pi}|^2\int dM_I^2\,\text{Im}\,[t_{\pi\pi,\pi\pi}^{I=0} (M_I)]
|\tilde t(M_I)|^2\,p_\pi^3(M_I) \ ,
\label{eq:gammafinal}
\end{align}
where the dependence on $M_{a_1}$ is explicitly highlighted.
Since the $a_1(1260)$ also has a large width we can do a convolution of the width of Eq.~\eqref{eq:gammafinal} with the mass distribution of the $a_1(1260)$ in analogy to Eq.~\eqref{eq:gammainter}. We use the most accurate results for the $a_1(1260)$ mass and width from the COMPASS collaboration~\cite{COMPASSa1}, 
\begin{eqnarray}
 &&M_{a_1}=(1255\pm 6^{+7}_{-17})~\mathrm{MeV}\ ,\nonumber\\
 &&\Gamma_{a_1}=(367\pm 9^{+28}_{-25})~\mathrm{MeV}\ ,\label{eq:a1p}
\end{eqnarray}
and obtain 
\begin{align}
\Gamma=\frac{1}{N}\int\limits_{(M_{a_1}-2\Gamma_{a_1})^2}^{(M_{a_1}+2\Gamma_{a_1})^2} d\tilde M_{a_1}^2\left(-\frac{1}{\pi}\right)\,\text{Im}\,\frac{1}{\tilde M_{a_1}^2-M_{a_1}^2+iM_{a_1}\Gamma_{a_1}}\,\Gamma(\tilde M_{a_1})\label{eq:convo}
\end{align}
with
\begin{align}
N=\int\limits_{(M_{a_1}-2\Gamma_{a_1})^2}^{(M_{a_1}+2\Gamma_{a_1})^2} d\tilde M_{a_1}^2\left(-\frac{1}{\pi}\right)\,\text{Im}\,\frac{1}{\tilde M_{a_1}^2-M_{a_1}^2+iM_{a_1}\Gamma_{a_1}} \ .
\end{align}


\section{Results}
\label{sec:results}
To calculate the $a_1\to\pi\sigma$ width we use the model of Ref.~\cite{npa} with a single channel $\ket{ \pi\pi,I=0}$ amplitude using unitary normalization (extra $1/\sqrt{2}$ in Eq. (\ref{eq:pipi})),
\begin{eqnarray}
 t^{I=0,u}_{\pi\pi,\pi\pi}=\frac{V^{I=0,u}}{1-V^{I=0,u}G} \ ,\label{eq:betheu}
\end{eqnarray}
with \begin{eqnarray}
  V^{I=0,u}=-\frac{1}{f^2}(M^2_{I}-\frac{m^2_\pi}{2}) \ ,   \label{eq:potu}
     \end{eqnarray}
 and $G$ the $\pi\pi$ loop function regularized with a cut off $q’_\mathrm{max}=750$ MeV, such as to get the $I=0$, $\pi\pi$ phase shifts. The coupling that we obtain is given in Eq.~\eqref{eq:cousigma}. Thus,
\begin{eqnarray}
 \vert g_{\sigma,\pi\pi}^{(u)}\vert=2.99~\mathrm{GeV};~\vert g_{\sigma,\pi\pi}\vert =\sqrt{2}\,g^{(u)}_{\sigma,\pi\pi}\ ,\label{eq:cousigma1}
 \end{eqnarray}and we have,
 \begin{eqnarray}
 t^{I=0}_{\pi\pi,\pi\pi}=2\,t^{I=0,(u)}_{\pi\pi,\pi\pi}\ .\label{eq:coupu}
\end{eqnarray}
We have also carried out the calculations using the model of Ref.~\cite{Liang}, considering explicitly the $\pi^+\pi^-,\pi^0\pi^0$, $K^+K^-,K^0\bar{K}^0$, and $\eta\eta$ channels, and using Eq. (\ref{eq:trel}), and the results are practically the same except that including the extra channels demands the use of a different cut off, $q_\mathrm{max}=600$ MeV. The results that we show are those based on Eqs. (\ref{eq:betheu})-(\ref{eq:coupu}). We shall estimate uncertainties at the end using results of different models for $\pi\pi$ scattering. 

\begin{figure}[tb]
 \begin{center}
  \includegraphics[scale=0.7]{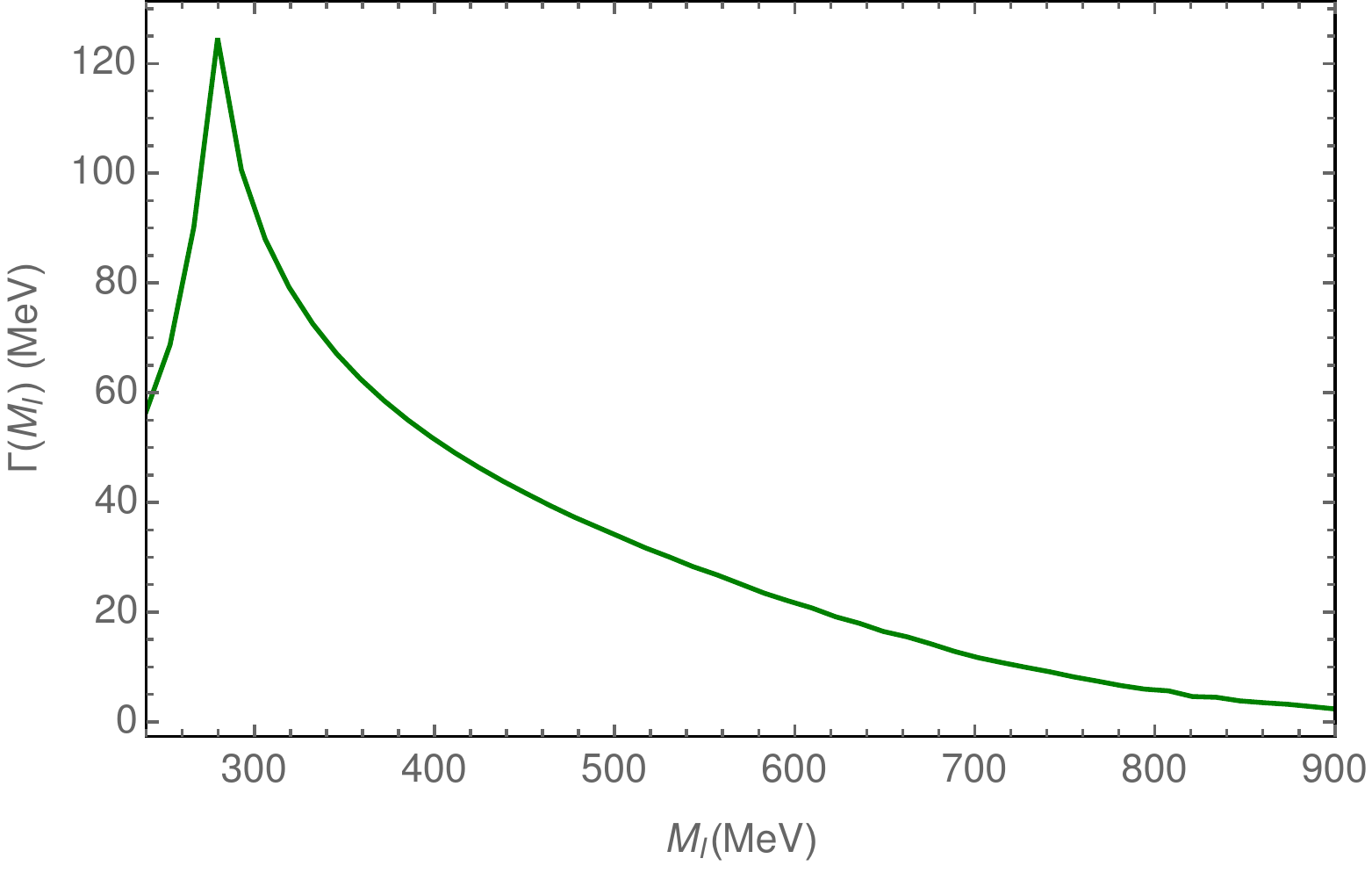}
 \end{center}
 \vspace{-0.7cm}
\caption{$\Gamma(a_1^+\to \pi^+\sigma)$ from Eq. \eqref{eq:width} as a function of $M_I$ for fixed $M_{a_1}=1255$ MeV.}
\label{fig:3gama}
\end{figure}

In Fig. \ref{fig:3gama} we show first the results obtained with Eq. (\ref{eq:width}) as a function of $M_I$. 
As expected for lower $\sigma$ masses $\Gamma$ grows since $p_\pi$ is bigger, and we find a cusp like peak at the two-pion threshold as it should be. We already see from Fig.~\ref{fig:3gama} that around $M_I\simeq 500$ MeV the width is of the order of $33$ MeV.

\begin{figure}[tb]
 \begin{center}
  \includegraphics[scale=0.7]{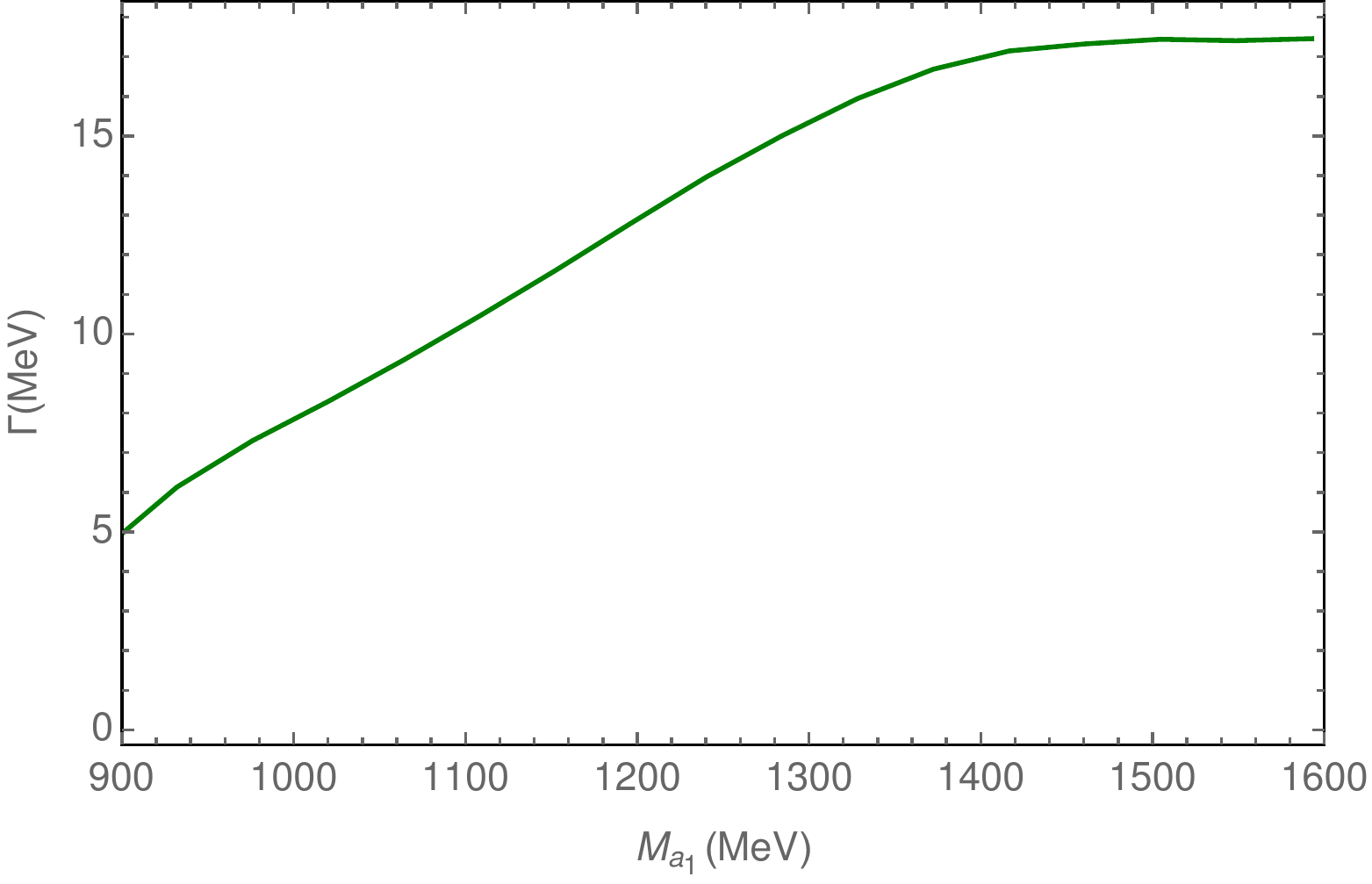}
 \end{center}
  \vspace{-0.8cm}
\caption{$\Gamma(a_1^+\to \pi^+\sigma)$ from Eq. \eqref{eq:gammafinal} as a function of $M_{a_1}$.}
\label{fig:4gma1}
\end{figure}

\begin{figure}[tb]
 \begin{center}
  \includegraphics[scale=0.7]{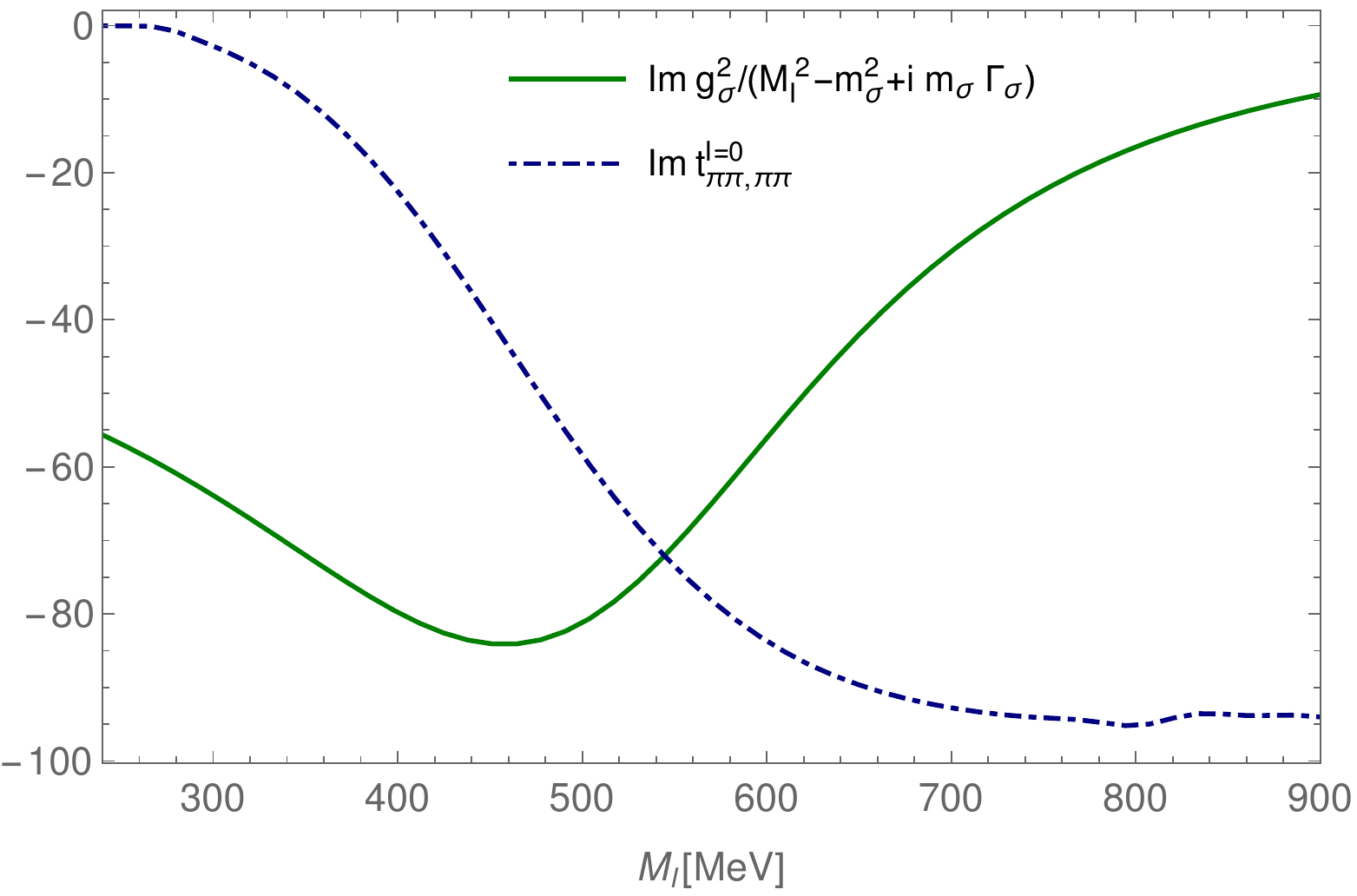}
 \end{center}
 \vspace{-0.8cm}
\caption{Comparison of the imaginary parts of $g^2_\sigma/(M^2_I-m^2_\sigma+i m_\sigma \Gamma_\sigma)$ and $t_{\pi\pi,\pi\pi}^{I=0}$.}
\label{fig:5compa}
\end{figure}

Next we use Eq. (\ref{eq:gammafinal}) in terms of $t^\mathrm{I=0}_{\pi\pi,\pi\pi}$ of Eqs. (\ref{eq:betheu})-(\ref{eq:coupu}). We show the results in terms of $M_{a_1}$ which we plot in Fig. \ref{fig:4gma1}. 
We see that $\Gamma$ has been reduced, since around $M_{a_1}=1255$ MeV it has a value about $14$ MeV, around one half what we estimated from Fig. \ref{fig:3gama}. We should explain this difference, which we do in Fig. \ref{fig:5compa}. 
There we plot the imaginary part of the Breit-Wigner form of the $\sigma$,
\begin{eqnarray}
 \mathrm{Im}\frac{\vert g_{\sigma,\pi\pi}\vert^2}{M^{2}_I-m^2_\sigma+im_\sigma\Gamma_\sigma}\ ,
\end{eqnarray}
versus that of the realistic $t^{I=0}_{\pi\pi,\pi\pi}$ amplitude, $\mathrm{Im}\,t^{I=0}_{\pi\pi,\pi\pi}$.
What we observe is that these two magnitudes are very different, both in strength and shape. In particular, the imaginary part of the full $\pi\pi$ amplitude goes to zero at threshold due to unitarity, as it should be, while the Breit-Wigner approximation does not. At higher $M_I$ the situation is opposite. This discrepancy should be stressed since in conventional experimental partial-wave analyses it is customary to use Breit-Wigner forms for the $\sigma$, although in some cases more sophisticated forms are taken, like in Ref.~\cite{newbes} based upon the formulation of Ref.~\cite{zoubugg}. Altogether we find about a factor of two difference in the evaluation of the width by both methods, the realistic one being the one obtained in terms of $t_{\pi\pi,\pi\pi}^{I=0}$, of course.

Finally, if we conduct the convolution of the result in Fig. \ref{fig:4gma1} with the $a_1(1260)$ mass distribution according to Eq.~(\ref{eq:convo}) we find 
\begin{equation}
 \Gamma= 13.4\,\mev\, .
 \label{eq:partialwidth}
\end{equation}
Compared with the $a_1(1260)$ width of Eq. (\ref{eq:a1p}), it gives a branching ratio of
\begin{equation}
    {\cal B}(a_1(1260)\to\pi\sigma)\simeq(3.3-4.0)\%
    \label{eq:br}
\end{equation}
considering the errors of Eq.~\eqref{eq:a1p}.
\subsection{Uncertainty analysis}
The branching ratio of Eq. (\ref{eq:br}) reflects only the uncertainties in the width of Eq. (\ref{eq:convo}). To estimate uncertainties from the use of other models we resort to a simple method by looking at the couplings obtained by different groups which are tabulated in Table 5 of Ref.~\cite{sigma}. We simply scale the result of Eq.~\eqref{eq:partialwidth} by the squared of the ratio of the couplings shown in that table. We show the results in Table \ref{tab:compa} and find a range of results from about $13$ MeV to $22$ MeV for the width of $a_1(1260)$ to $\sigma\pi$.
\begin{table}[tb]
\begin{center}
{\renewcommand{\arraystretch}{1.4}
\setlength\tabcolsep{0.4cm}
 \begin{tabular}{|ccc|} \hline
  Ref.&$g_{\sigma\pi\pi}(\mathrm{GeV})$&$\Gamma(\mathrm{MeV})$\\\hline
  \cite{Caprini}&$3.58\pm 0.03$&$19.3$\\
  \cite{Pelaezrios}&$3.5$&$18.3$\\
  \cite{Masjuan,Caprini2}&$3.8\pm 0.4$&$21.6$\\
  This work / \cite{npa} &$2.99$&$13.4$\\\hline
 \end{tabular}}
  \caption{Estimation of the uncertainties in $\Gamma$ from the coupling $g_{\sigma,\pi\pi}$ with some $g_{\sigma,\pi\pi}$ taken from Ref.~\cite{sigma}.}
\label{tab:compa}
\end{center}
\end{table}
The branching ratios obtained, $(3.3-6.6)$\%, are in the range of some of the small values quoted in Eqs. (\ref{eq:exp}), but much smaller than the rate quoted in Ref.~\cite{Asner}, Eq. (\ref{eq:asner}). We should certainly note the large dispersion of experimental results in Eqs. (\ref{eq:asner}), (\ref{eq:exp}) that might come from treating the $\sigma$ as an ordinary resonance as discussed before. The formulation presented here should help perform more accurate analyses of data in future analyses.
Actually, with this work finished, there has been a recent experiment \cite{newbes} for the $D^+_s \to \pi^+ \pi^- \pi^+ \eta$ decay with emphasis on different partial wave contributions, and from where one can induce the branching ratio that we have evaluated.
From Table I of that work one finds a weight for $a_1(1260)^+ \eta$, with $a_1^+$ observed in $\rho^0 \pi^+$, $W_1=55.4 \pm 3.9 \pm 2.0$.
Then $a_1^+ \eta$, with $a_1^+$ observed in $f_0(500) \pi^+$,
has a weight of $W_2=8.1 \pm 1.9 \pm 2.1$.
But this only accounts for $f_0(500)\to\pi^+\pi^-$. The $\pi^0\pi^0$ decay has a strength of $\nicefrac{1}{2}$ of that of $\pi^+\pi^-$. Furthermore,
since the weight for  $a_1(1260)$ decay into $\rho^+ \pi^0$ is the same as for $\rho^0 \pi^+$, and the total branching ratio of $a_1 \to \rho \pi$ is $60\%$~\cite{PDG},
we have 
\begin{equation}
 \dfrac{\Gamma(a_1 \to \sigma \pi)}{\Gamma_{a_1}}= \dfrac{3W_2/2}{2W_1/0.6}=(6.6\pm 2.4)\%,
\label{eq:besBR}
\end{equation}
where we have summed errors in quadrature.
This number is in good agreement with our results. It is interesting to see that with our $a_1\to\rho\pi$ coupling of Eq.~\eqref{eq:ga1}, we obtain  a branching fraction of $a_1$ to $\rho\pi$ of 51\%, close to the one reported in the PDG~\cite{PDG}. If, instead of the rather uncertain experimental $a_1\to\rho\pi$ 60\% branching ratio, we use our theoretical value of 51\%, the numbers in Eq.~\eqref{eq:besBR} change to $(5.6\pm 2)\%$. Hence, a range of $\Gamma(a_1 \to \sigma \pi)/\Gamma_{a_1}\approx (3.6-9)\%$ is a safe estimate from the experiment of Ref.~\cite{newbes}. 

\section{Conclusions}
    We have evaluated the decay width of the $a_1(1260)$ resonance into $\pi \sigma (f_0(500))$ taking into account that the $a_1(1260)$ is a state dynamically generated by the vector-pseudoscalar interaction, concretely with the channels $\rho \pi$ and $K^* \bar K$, and the $\sigma$ arises from the unitarized interaction of pseudoscalar-pseudoscalar mesons, mostly of two pions. Given this starting point, the mechanism that leads to the $a_1\to\pi\sigma$ decay is given by a triangle mechanism in which the $a_1(1260)$ decays to $\rho \pi$, the $\rho$ decays to $\pi \pi$ and one of these pions, together with the one coming from the $a_1 \to \rho \pi$ decay, fuse into the $\sigma$ meson.  The mechanism allows to have the intermediate particles on shell, but does not develop a triangle singularity, yet it provides a relatively large strength for that decay. We find decay widths in this channel of the order of $13-22\, \mev$, which amount to a branching fraction of $(3.3-6.6)$\%. This is a small fraction and several experiments, so far largely contradictory, have been devoted to estimate that width.
    On the other hand, the recent BESIII experiment \cite{newbes} allows to extract this branching ratio to be in the range $(3.6-9)\%$,
    which overlaps with the range that we obtain.
  
 One of the findings is that the width obtained depends much on how the mass distribution  of the $\sigma$ is considered.  We found that treating the $\sigma$ as an ordinary Breit-Wigner resonance, using  realistic coupling, mass and width obtained in theoretical works, gives rise to results which differ substantially from those obtained by using the imaginary part of the $I=0$, $\pi \pi$ scattering  amplitude as the mass distribution. This should serve as a warning for future experimental analyses of this partial decay width.
 The work of Ref.~\cite{newbes} uses a more elaborate and realistic picture for the $\sigma$ resonance extracted from Ref.~\cite{zoubugg}.

Any information on the $\pi\sigma$ and other subdominant branching ratios of the $a_1(1260)$ helps assess systematic effects in recent efforts to calculate the properties of three-body resonances in lattice QCD~\cite{Mai:2021nul}. In particular, the smallness of that branching ratio found in this study strengthens the truncation to $\pi\rho$ channels made in the finite-volume frameworks of Refs.~\cite{RocaFinite, Mai:2021nul}.

     The results of this study are tied to the assumptions about the nature of both the $a_1(1260)$ and the $\sigma$ resonances as being dynamically generated from the pseudoscalar-vector and pseudoscalar-pseudoscalar interaction. Given the ongoing discussion on the nature of the axial vector and scalar meson resonances, the precise determination of this width should provide valid information concerning this relevant issue, and it is our hope that the present work stimulates further efforts in this direction.

\begin{acknowledgments}
R. M. acknowledges support from the CIDEGENT program with Ref. CIDEGENT/2019/015 and from the spanish national grants PID2019-106080GB-C21 and PID2020-112777GB-I00.
This work is  partly supported by the National Natural Science Foundation of China under Grants No. 11975083 and No. 11947413.
This work is also partly supported by the Spanish Ministerio de Economia y Competitividad
and European FEDER funds under Contracts No. FIS2017-84038-C2-1-P B
and by Generalitat Valenciana under contract PROMETEO/2020/023.
This project has received funding from the European Unions Horizon 2020 research and innovation programme
under grant agreement No. 824093 for the ``STRONG-2020" project. 
This material is based upon work supported by the National Science Foundation under Grant No. PHY-2012289 and the U.S. Department of Energy under Award Number DE-SC0016582 and DE-AC05-06OR23177. MD is grateful for the hospitality of the University of Valencia, where part of this work was done.
\end{acknowledgments}

\appendix

\section{Coupling of $\sigma$ to $\pi\pi$}
\label{sec:appa}
\begin{figure}[htb]
    \centering
    \includegraphics[scale=0.6]{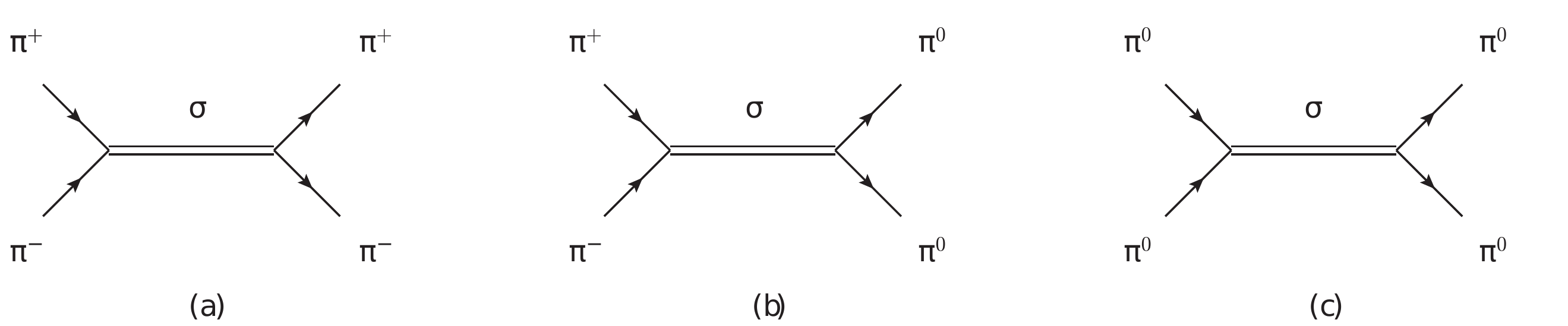}
    \caption{Amplitudes $\pi\pi\to\pi\pi$ through an isoscalar source.}
    \label{fig:twopi}
\end{figure}
We consider here the couplings of $\sigma$ to $\pi^+\pi^-,\pi^0\pi^0$ taking into account the identity of the $\pi^0$. An isoscalar coupling to $\pi\pi$ is given by a Lagrangian of the type,
\begin{eqnarray}
 {\cal L}=\alpha\, \vec{\phi}\cdot\vec{\phi}=\alpha\,(-1)^\mu\phi_\mu \phi_{-\mu},\quad \mu=-1,0,+1
 \label{eq:lag}
\end{eqnarray}
with 
\begin{eqnarray}
&& \phi_{-1}=\frac{1}{\sqrt{2}}(\phi_1+i\phi_2),\nonumber\\
 &&\phi_{0}=\phi_3,\nonumber\\
 &&\phi_{+1}=-\frac{1}{\sqrt{2}}(\phi_1-i\phi_2),
\end{eqnarray}
where $\phi_{-}$ destroys a $\pi^-$, creates a $\pi^+$, while $\phi_+$ destroys a $\pi^+$ and creates a $\pi^-$. 

Using the Lagrangians of Eq. (\ref{eq:lag}) and the two ways to destroy or create $\pi^+\pi^-$, or $\pi^0\pi^0$, we obtain
\begin{eqnarray}
 t^{I=0}_{\pi^+\pi^-,\pi^+\pi^-}=t^{I=0}_{\pi^+\pi^-,\pi^0\pi^0}=t^{I=0}_{\pi^0\pi^0,\pi^0\pi^0}=\frac{(2\alpha)^2}{s-m^2_\sigma},\label{eq:ide}
\end{eqnarray}
and using the $\pi\pi (I=0)$ states of Eq. (\ref{eq:pipi}) we find
\begin{eqnarray}
 &&t^{I=0}_{\pi\pi,\pi\pi}=\frac{1}{3}(4t^{I=0}_{\pi^+\pi^-,\pi^+\pi^-}+4t^{I=0}_{\pi^+\pi^-,\pi^0\pi^0}+t^{I=0}_{\pi^0\pi^0,\pi^0\pi^0})\nonumber\\
 &&=\frac{9}{3}t^{I=0}_{\pi^+\pi^-,\pi^0\pi^0}=\frac{9}{3}\frac{(2\alpha)^2}{s-m^2_\sigma}\equiv\frac{g^2_{\sigma,\pi\pi}}{s-m^2_\sigma}.\label{eq:cou}
\end{eqnarray}
Hence we find:
\begin{eqnarray}
g_{\sigma,\pi^+\pi^-}\equiv g_{\sigma,\pi^0\pi^0}=2\alpha=\frac{1}{\sqrt{3}}g_{\sigma,\pi\pi}.
\end{eqnarray}


\section{Boosted momenta}
\label{sec:appb}
The vector $\vec q\,'$ defined in the rest frame of the $\sigma$ is given by
\begin{eqnarray}
 \vec{q}\,'=\left[\left(\frac{E_\sigma}{M_{I}}-1\right)\frac{\vec{q}\cdot \vec{p}_\sigma}{\vec{p}\,^2_\sigma}-\frac{q^0}{M_I}\right]\vec{p}_\sigma+\vec{q},\label{eq:boost}
\end{eqnarray}
where
\begin{eqnarray}
 &&\vec{p}_\sigma=-\vec{p}_\pi,\nonumber\\
 &&q^0=\omega(\vec q),\nonumber\\
 &&E_\sigma=M_{a_1}-\omega(\vec p_\pi),\nonumber\\
 &&|\vec{p}_\pi|=\frac{\lambda^{1/2}(M^2_{a_1},m^2_\pi,M^{2}_I)}{2M_{a_1}},
\end{eqnarray}
and $M_I$ is the invariant mass carried by the $\sigma$ in the diagrams of Figs.~\ref{fig:triangles}(a), \ref{fig:details}, and \ref{fig:twopi}.



\begin{thebibliography}{}

\bibitem{Isgur}
  S.~Godfrey and N.~Isgur,
  Phys.\ Rev.\ D {\bf 32}, 189 (1985).

\bibitem{Vijande}
  J.~Vijande, F.~Fernandez and A.~Valcarce,
  J.\ Phys.\ G {\bf 31}, 481 (2005).

\bibitem{Lutz}
  M.~F.~M.~Lutz and E.~E.~Kolomeitsev,
  Nucl.\ Phys.\ A {\bf 730}, 392 (2004).

\bibitem{Luis}
  L.~Roca, E.~Oset and J.~Singh,
  Phys.\ Rev.\ D {\bf 72}, 014002 (2005).

\bibitem{GengRoca}
  L.~S.~Geng, E.~Oset, L.~Roca and J.~A.~Oller,
  Phys.\ Rev.\ D {\bf 75}, 014017 (2007).

\bibitem{RenGeng}
  Y.~Zhou, X.~L.~Ren, H.~X.~Chen and L.~S.~Geng,
  Phys.\ Rev.\ D {\bf 90}, no. 1, 014020 (2014).

\bibitem{Birse}
  M.~C.~Birse,
  Z.\ Phys.\ A {\bf 355}, 231 (1996).

\bibitem{Nagahiro}
  H.~Nagahiro, L.~Roca, A.~Hosaka and E.~Oset,
  Phys.\ Rev.\ D {\bf 79}, 014015 (2009).

\bibitem{Jorgivan}
J.~M.~Dias, G.~Toledo, L.~Roca and E.~Oset,
  Phys.\ Rev.\ D {\bf 103}, 116019 (2021).

\bibitem{hidden1}
  M.~Bando, T.~Kugo and K.~Yamawaki,
  Phys.\ Rept.\  {\bf 164}, 217 (1988).

\bibitem{hidden2}
  M.~Harada and K.~Yamawaki,
  Phys.\ Rept.\  {\bf 381}, 1 (2003).

\bibitem{hidden4}
  U.~G.~Meissner,
  Phys.\ Rept.\  {\bf 161}, 213 (1988).

\bibitem{Rafael}
  G.~Ecker, J.~Gasser, H.~Leutwyler, A.~Pich and E.~de Rafael,
  Phys.\ Lett.\ B {\bf 223}, 425 (1989).

\bibitem{Nakamura}
  S.~X.~Nakamura,
  Phys.\ Rev.\ D {\bf 93}, no. 1, 014005 (2016).

\bibitem{Sasa}
S.~Prelovsek, C.~B.~Lang, D.~Mohler and M.~Vidmar,
PoS \textbf{LATTICE2011}, 137 (2011).

\bibitem{RocaFinite}
  L.~Roca and E.~Oset,
  Phys.\ Rev.\ D {\bf 85}, 054507 (2012).

\bibitem{Lang}
  C.~B.~Lang, L.~Leskovec, D.~Mohler and S.~Prelovsek,
  JHEP {\bf 1404}, 162 (2014).

\bibitem{Dudek}
  A.~J.~Woss, C.~E.~Thomas, J.~J.~Dudek, R.~G.~Edwards and D.~J.~Wilson,
  Phys.\ Rev.\ D {\bf 100}, no. 5, 054506 (2019).

\bibitem{Mai:2021nul}
M.~Mai, A.~Alexandru, R.~Brett, C.~Culver, M.~D{\"o}ring, F.~X.~Lee and D.~Sadasivan,
[arXiv:2107.03973 [hep-lat]].

\bibitem{Mai:2017vot}
M.~Mai, B.~Hu, M.~Doring, A.~Pilloni and A.~Szczepaniak,
Eur. Phys. J. A \textbf{53},  177 (2017)
[arXiv:1706.06118 [nucl-th]].

\bibitem{Mai:2017bge}
M.~Mai and M.~D\"oring,
Eur. Phys. J. A \textbf{53}, no.12, 240 (2017)
[arXiv:1709.08222 [hep-lat]].

\bibitem{Aceti}
  F.~Aceti, J.~M.~Dias and E.~Oset,
  Eur.\ Phys.\ J.\ A {\bf 51}, 48 (2015).

\bibitem{AcetiKK}
  F.~Aceti, J.~J.~Xie and E.~Oset,
  Phys.\ Lett.\ B {\bf 750}, 609 (2015).

\bibitem{Zou}
  J.~J.~Wu, X.~H.~Liu, Q.~Zhao and B.~S.~Zou,
  Phys.\ Rev.\ Lett.\  {\bf 108}, 081803 (2012).

\bibitem{Acetieta}
  F.~Aceti, W.~H.~Liang, E.~Oset, J.~J.~Wu and B.~S.~Zou,
  Phys.\ Rev.\ D {\bf 86}, 114007 (2012).

\bibitem{BESeta}
  M.~Ablikim {\it et al.} [BESIII Collaboration],
  Phys.\ Rev.\ Lett.\  {\bf 108}, 182001 (2012).

\bibitem{Achasov}
  N.~N.~Achasov, A.~A.~Kozhevnikov and G.~N.~Shestakov,
  Phys.\ Rev.\ D {\bf 92}, 036003 (2015).

\bibitem{QiangZhao}
  M.~C.~Du and Q.~Zhao,
  Phys.\ Rev.\ D {\bf 100}, no. 3, 036005 (2019).

\bibitem{He}
  Y.~Huang, C.~j.~Xiao, Q.~F.~L., R.~Wang, J.~He and L.~Geng,
  Phys.\ Rev.\ D {\bf 97}, no. 9, 094013 (2018).

\bibitem{HuangGeng}
  Y.~Huang, M.~Z.~Liu, J.~X.~Lu, J.~J.~Xie and L.~S.~Geng,
  Phys.\ Rev.\ D {\bf 98}, no. 7, 076012 (2018).

\bibitem{XieChi}
  J.~J.~Xie, G.~Li and X.~H.~Liu,
  Chin.\ Phys.\ C {\bf 44}, no. 11, 114104 (2020).

\bibitem{Xiemore}
  Z.~M.~Ding, H.~Y.~Jiang and J.~He,
  Eur.\ Phys.\ J.\ C {\bf 80}, no. 12, 1179 (2020).

\bibitem{HuangXie}
  Y.~Huang, J.~X.~Lu, J.~J.~Xie and L.~S.~Geng,
  Eur.\ Phys.\ J.\ C {\bf 80}, no. 10, 973 (2020).

\bibitem{YuKe}
  Y.~K.~Hsiao, Y.~Yu and B.~C.~Ke,
  Eur.\ Phys.\ J.\ C {\bf 80}, no. 9, 895 (2020).

\bibitem{GengXie}
  X.~Z.~Ling, M.~Z.~Liu, J.~X.~Lu, L.~S.~Geng and J.~J.~Xie,
  arXiv:2102.05349 [hep-ph].

\bibitem{Karplus}
  R.~Karplus, C.~M.~Sommerfield and E.~H.~Wichmann,
  Phys.\ Rev.\  {\bf 111}, 1187 (1958).

\bibitem{Landau}
  L.~D.~Landau,
  Nucl.\ Phys.\  {\bf 13}, no. 1, 181 (1959)
  [Sov.\ Phys.\ JETP {\bf 10}, no. 1, 45 (1960)]
  [Zh.\ Eksp.\ Teor.\ Fiz.\  {\bf 37}, no. 1, 62 (1959)].

\bibitem{Norton}
  S.~Coleman and R.~E.~Norton,
  Nuovo Cim.\  {\bf 38}, 438 (1965).

\bibitem{BayarGuo}
  M.~Bayar, F.~Aceti, F.~K.~Guo and E.~Oset,
  Phys.\ Rev.\ D {\bf 94}, no. 7, 074039 (2016).

\bibitem{Guorev}
  F.~K.~Guo, X.~H.~Liu and S.~Sakai,
  Prog.\ Part.\ Nucl.\ Phys.\  {\bf 112}, 103757 (2020).

\bibitem{COMPASS}
  C.~Adolph {\it et al.} [COMPASS Collaboration],
  Phys.\ Rev.\ Lett.\  {\bf 115}, no. 8, 082001 (2015).

\bibitem{Oka}
  X.~H.~Liu, M.~Oka and Q.~Zhao,
  Phys.\ Lett.\ B {\bf 753}, 297 (2016).

\bibitem{Mikha}
  M.~Mikhasenko, B.~Ketzer and A.~Sarantsev,
  Phys.\ Rev.\ D {\bf 91}, no. 9, 094015 (2015).

\bibitem{AceDai}
  F.~Aceti, L.~R.~Dai and E.~Oset,
  Phys.\ Rev.\ D {\bf 94}, no. 9, 096015 (2016).

\bibitem{Alexeev}
  M.~G.~Alexeev {\it et al.} [COMPASS Collaboration],
  arXiv:2006.05342 [hep-ph].

\bibitem{IkenoRaq}
  N.~Ikeno, R.~Molina and E.~Oset,
  arXiv:2103.01712 [nucl-th].

\bibitem{RaquelIke}
  R.~Molina, N.~Ikeno and E.~Oset,
  arXiv:2102.05575 [nucl-th].

\bibitem{dibaryon}
  M.~Bashkanov {\it et al.},
  Phys.\ Rev.\ Lett.\  {\bf 102}, 052301 (2009).

\bibitem{Misha}
  D.~Sadasivan, M.~Mai, H.~Akdag and M.~D\"oring,
  Phys.\ Rev.\ D {\bf 101}, no. 9, 094018 (2020);
  Erratum: [Phys.\ Rev.\ D {\bf 103}, no. 1, 019901 (2021)].

\bibitem{ZhangXie}
  X.~Zhang and J.~J.~Xie,
  Commun.\ Theor.\ Phys.\  {\bf 70}, no. 1, 060 (2018).

\bibitem{DaiRoca}
L.~R.~Dai, L.~Roca and E.~Oset,
Eur. Phys. J. C \textbf{80}, no.7, 673 (2020).

\bibitem{PDG}
 P.~A.~Zyla {\it et al.} [Particle Data Group],
 Prog. Theor. Exp. Phys. {\bf 2020}, no. 8, 083C01 (2020) and 2021 update.

\bibitem{Asner}
  D.~M.~Asner {\it et al.} [CLEO Collaboration],
  Phys.\ Rev.\ D {\bf 61}, 012002 (2000).

\bibitem{Salvini}
  P.~Salvini {\it et al.} [OBELIX Collaboration],
  Eur.\ Phys.\ J.\ C {\bf 35}, 21 (2004).

\bibitem{Akhme}
  R.~R.~Akhmetshin {\it et al.} [CMD-2 Collaboration],
  Phys.\ Lett.\ B {\bf 466}, 392 (1999).

\bibitem{Longa}
  R.~S.~Longacre,
  Phys.\ Rev.\ D {\bf 26}, 82 (1982).


\bibitem{Chung}
  S.~U.~Chung {\it et al.},
  Phys.\ Rev.\ D {\bf 65}, 072001 (2002).


\bibitem{Colangelo}
  G.~Colangelo, J.~Gasser and H.~Leutwyler,
  Nucl.\ Phys.\ B {\bf 603}, 125 (2001).


\bibitem{sigma}
  J.~R.~Pelaez,
  Phys.\ Rept.\  {\bf 658}, 1 (2016).


\bibitem{Kamano:2011ih}
H.~Kamano, S.~X.~Nakamura, T.~S.~H.~Lee and T.~Sato,
Phys. Rev. D \textbf{84}, 114019 (2011)
[arXiv:1106.4523 [hep-ph]].


\bibitem{Janssen:1993nj}
G.~Janssen, J.~W.~Durso, K.~Holinde, B.~C.~Pearce and J.~Speth,
Phys. Rev. Lett. \textbf{71}, 1978-1981 (1993).


\bibitem{Wagner:2007wy}
M.~Wagner and S.~Leupold,
Phys. Lett. B \textbf{670}, 22-26 (2008)
[arXiv:0708.2223 [hep-ph]].

\bibitem{Wagner:2008gz}
M.~Wagner and S.~Leupold,
Phys. Rev. D \textbf{78}, 053001 (2008)
[arXiv:0801.0814 [hep-ph]].


\bibitem{ALEPH:2005qgp}
S.~Schael \textit{et al.} [ALEPH],
Phys. Rept. \textbf{421}, 191-284 (2005)
[arXiv:hep-ex/0506072 [hep-ex]].


\bibitem{npa}
  J.~A.~Oller and E.~Oset,
  Nucl.\ Phys.\ A {\bf 620}, 438 (1997);
  Erratum: [Nucl.\ Phys.\ A {\bf 652}, 407 (1999)].

\bibitem{ARGUS}
  H.~Albrecht {\it et al.} [ARGUS Collaboration],
  Z.\ Phys.\ C {\bf 58}, 61 (1993).

\bibitem{nsd}
J.~A.~Oller and E.~Oset,
Phys. Rev. D \textbf{60}, 074023 (1999).

\bibitem{SakaiRamos}
S.~Sakai, E.~Oset and A.~Ramos,
Eur. Phys. J. A \textbf{54}, 10 (2018).

\bibitem{DaniJuan}
D.~Gamermann, J.~Nieves, E.~Oset and E.~Ruiz Arriola,
Phys. Rev. D \textbf{81}, 014029 (2010).

\bibitem{Liang}
W.~H.~Liang and E.~Oset,
Phys. Lett. B \textbf{737}, 70-74 (2014).

\bibitem{COMPASSa1}
M.~Alekseev \textit{et al.} [COMPASS],
Phys. Rev. Lett. \textbf{104}, 241803 (2010).
\bibitem{newbes}
M.~Ablikim \textit{et al.} [BESIII],
[arXiv:2106.13536 [hep-ex]].
\bibitem{zoubugg}
D.~V.~Bugg, B.~S.~Zou and A.~V.~Sarantsev,
Nucl. Phys. B \textbf{471}, 59-89 (1996).
\bibitem{Caprini}
I.~Caprini,
Phys. Rev. D \textbf{77}, 114019 (2008).
\bibitem{Pelaezrios}
J.~R.~Pelaez and G.~Rios,
Phys. Rev. D \textbf{82}, 114002 (2010).
\bibitem{Masjuan}
P.~Masjuan, J.~Ruiz de Elvira and J.~J.~Sanz-Cillero,
Phys. Rev. D \textbf{90}, no.9, 097901 (2014).
\bibitem{Caprini2}
I.~Caprini, P.~Masjuan, J.~Ruiz de Elvira and J.~J.~Sanz-Cillero,
Phys. Rev. D \textbf{93}, no.7, 076004 (2016).

\end{thebibliography}
\end{document}